\documentclass[12pt]{article}
\usepackage{epsfig}
\usepackage{amssymb}
\usepackage{amsmath}
\usepackage{amsfonts}
\usepackage{graphicx}
\usepackage{mathrsfs}
\usepackage[dvips]{color}
\usepackage{multirow}

\newcommand{\bsigma}{\boldsymbol{\sigma}}

\newcommand{\R}{\mathbb{R}}
\newcommand{\C}{\mathbb{C}}

\newcommand{\ff}{\mathfrak{f}}
\newcommand{\fg}{\mathfrak{g}}

\newcommand{\fz}{\mathfrak{z}}

\newcommand{\fK}{\mathfrak{K}}

\newcommand{\fM}{\mathfrak{M}}

\newcommand{\bfe}{\mathbf{e}}

\newcommand{\bk}{\mathbf{k}}

\newcommand{\bfr}{\mathbf{r}}
\newcommand{\bp}{\mathbf{p}}

\newcommand{\bI}{\mathbf{I}}

\newcommand{\bcL}{\boldsymbol{\cL}}
\newcommand{\bM}{\mathbf{M}}

\newcommand{\bS}{\mathbf{S}}

\newcommand{\cH}{\mathcal{H}}

\newcommand{\cF}{\mathcal{F}}

\newcommand{\cK}{\mathcal{K}}
\newcommand{\cL}{\mathcal{L}}

\newcommand{\cP}{\mathcal{P}}

\newcommand{\cR}{\mathcal{R}}
\newcommand{\cS}{\mathcal{S}}
\newcommand{\cT}{\mathcal{T}}
\newcommand{\cU}{\mathcal{U}}

\newcommand{\be}{\begin{equation}}
\newcommand{\ee}{\end{equation}}
\newcommand{\bea}{\begin{eqnarray}}
\newcommand{\eea}{\end{eqnarray}}
\newcommand{\nn}{\nonumber}
\newcommand{\kt}{\rangle}
\newcommand{\br}{\langle}

\newcommand{\ed}{\end{document}}

\newcommand{\bi}{\begin{itemize}}
\newcommand{\ei}{\end{itemize}}

\newcommand{\bce}{\begin{center}}
\newcommand{\ece}{\end{center}}
\newcommand{\sA}{\mathscr{A}}
\newcommand{\sB}{\mathscr{B}}

\newcommand{\sF}{\mathscr{F}}

\newcommand{\sV}{\mathscr{V}}

\newcommand{\bPi}{{\boldsymbol{\Pi}}}

\newcommand{\bcK}{{\boldsymbol{\cK}}}

\newcommand{\bfM}{{\boldsymbol{\fM}}}
\newcommand{\bcH}{{\boldsymbol{\cH}}}
\newcommand{\bcU}{{\boldsymbol{\cU}}}

\newcommand{\bzero}{{\boldsymbol{0}}}

\newcommand{\for}{{\mbox{\rm for}}}

\begin{document}

\title{Propagating-wave approximation in two-dimensional potential scattering}

%\title{Nonlocal potentials that do not couple to evanescent waves and a pseudo-classical approximation for potential\\ scattering in two dimensions}

%\title{Potentials that do not couple to evanescent waves and pseudo-classical scattering in two dimensions}

\author{Farhang Loran\thanks{E-mail address: loran@iut.ac.ir}~ and
Ali~Mostafazadeh\thanks{E-mail address:
amostafazadeh@ku.edu.tr}\\[6pt]
$^{*}$Department of Physics, Isfahan University of Technology, \\ Isfahan 84156-83111, Iran\\[6pt]
$^\dagger$Departments of Mathematics and Physics, Ko\c{c}
University,\\  34450 Sar{\i}yer, Istanbul, Turkey}

\date{ }
\maketitle

\begin{abstract}
We introduce a nonperturbative approximation scheme for performing scattering calculations in two dimensions that involves neglecting the contribution of the evanescent waves to the scattering amplitude. This corresponds to replacing the interaction potential $v$ with an associated energy-dependent nonlocal potential ${\mathscr{V}}_k$ that does not couple to the evanescent waves. The scattering solutions $\psi(\mathbf{r})$ of the Schr\"odinger equation, $(-\nabla^2+{\mathscr{V}}_k)\psi(\mathbf{r})=k^2\psi(\mathbf{r})$, has the remarkable property that their Fourier transform $\tilde\psi(\mathbf{p})$ vanishes unless $\mathbf{p}$ corresponds to the momentum of a classical particle whose magnitude equals $k$. We construct a transfer matrix for this class of nonlocal potentials and explore its representation in terms of the evolution operator for an effective non-unitary quantum system. We show that the above approximation reduces to the first Born approximation for weak potentials, and similarly to the semiclassical approximation, becomes valid at high energies. Furthermore, we identify an infinite class of complex potentials for which this approximation scheme is exact. We also discuss the appealing practical and mathematical aspects of this scheme.

%\vspace{2mm}

%\noindent PACS numbers: 03.65.Nk, 42.25.Bs\vspace{2mm}

%\noindent Keywords: Exceptional point, pseudo-Hermitian operator, scattering, transfer matrix, biorthonormal system
\end{abstract}

\section{Introduction}
\label{S1}

Evanescent waves are time-harmonic waves, $e^{-i\omega t}\psi(\bfr)$, that undergo exponential damping or growth in regions of space where the interaction ceases to exist. They arise in space dimensions higher than one and are responsible for the major differences between the behavior of waves propagating in one dimension and those propagating in two and higher dimensions. Suppose that $\psi$ solves the stationary Schr\"odinger equation,
	\be
	[-\partial_x^2-\partial_y^2+v(x,y)]\psi(x,y)=k^2\psi(x,y),~~~(x,y)\in\R^2,
	\label{sch-eq}
	\ee
in two dimensions, where $v$ is a real or complex scattering potential, $k$ is the wavenumber, and we use units where $\hbar=2m=1$. In regions $\cR$ where the potential vanishes, (\ref{sch-eq}) reduces to the Helmholtz equation, $(\nabla^2+k^2)\psi=0$, whose solutions are superpositions of the plane-wave solutions, 
	\be
	e^{\pm i\sqrt{k^2-p^2}\,x}e^{ipy}~~{\rm with}~~p\in(-k,k),
	\label{PW}
	\ee
and the evanescent-wave solutions,
	\be
	e^{\pm\sqrt{p^2-k^2}\,x}e^{ipy}~~{\rm with}~~p\notin(-k,k).
	\label{EW}
	\ee	 
We can express the plane-wave solutions (\ref{PW}) in the familiar form $e^{i\bp\cdot\bfr}$, where $\bp$ is the real wave vector having the $x$- and $y$-components, $p_x:=\pm\sqrt{k^2-p^2}$ and $p_y:=p$. Therefore, we can identify $\bp$ with the momentum of a classical particle. It is clear that this correspondence does not extend to the evanescent-wave solutions~(\ref{EW}); they do not have a classical counterpart. The purpose of the present article is to explore the consequences of ignoring the contribution of the evanescent waves (\ref{EW}) to the scattering features of the potential. This corresponds to a particular ``quasi-classical approximation'' scheme which is not to be confused with the standard semiclassical (WKB) approximation \cite{ford-1959,berry-1972,koeling-1975,adhikari-2008}. To avoid possible confusing we call it the  ``propagating-wave approximation.'' 

If $v$ is a short-range potential \cite{yafaev}, i.e., $r v(\bfr)\to 0$ as $r\to\infty$, the Schr\"odinger equation (\ref{sch-eq}) admits scattering solutions satisfying 
	\be
	\psi(\bfr)\to
	\frac{1}{2\pi}\Big[e^{i\bk_0\cdot\bfr}+\sqrt{\frac{i}{kr}}\,e^{ikr}\ff(\theta) \Big]
	~~\for~~r\to\infty,
	\label{scattering}
	\ee
where $\bfr:=x\, \bfe_x+y\,\bfe_y$, $\bfe_u$ is the unit vector along the $u$-axis for $u\in\{x,y\}$, $\bk_0\in\R^2$ is the incident wave vector,  $(r,\theta)$ are the polar coordinates of $\bfr$, and $\ff(\theta)$ is the scattering amplitude. Note also that $|\bk_0|=k$ and $\bk_0\cdot\bfr:=kr\cos(\theta-\theta_0)$, where $\theta_0$ is the incident angle which specifies the direction of $\bk_0$ according to $\bk_0/k=\cos\theta_0\bfe_x+\sin\theta_0\bfe_y$. %The solution of the scattering problem for $v$ means the determination of the scattering amplitude $\ff(\theta)$.

\begin{figure} 
        \begin{center}
        \includegraphics[scale=.23]{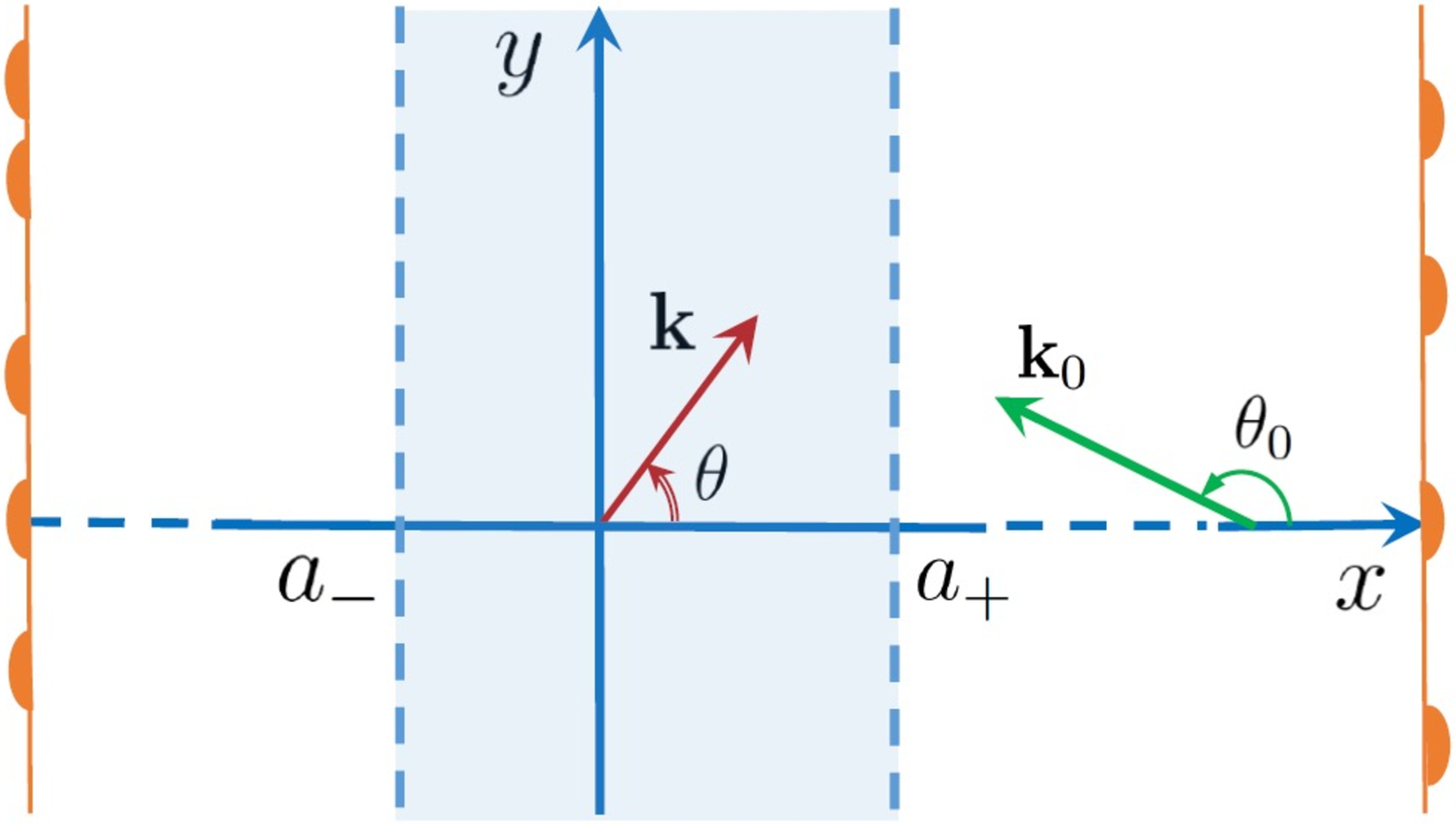}~~~~~~~~~~~~
        \includegraphics[scale=.23]{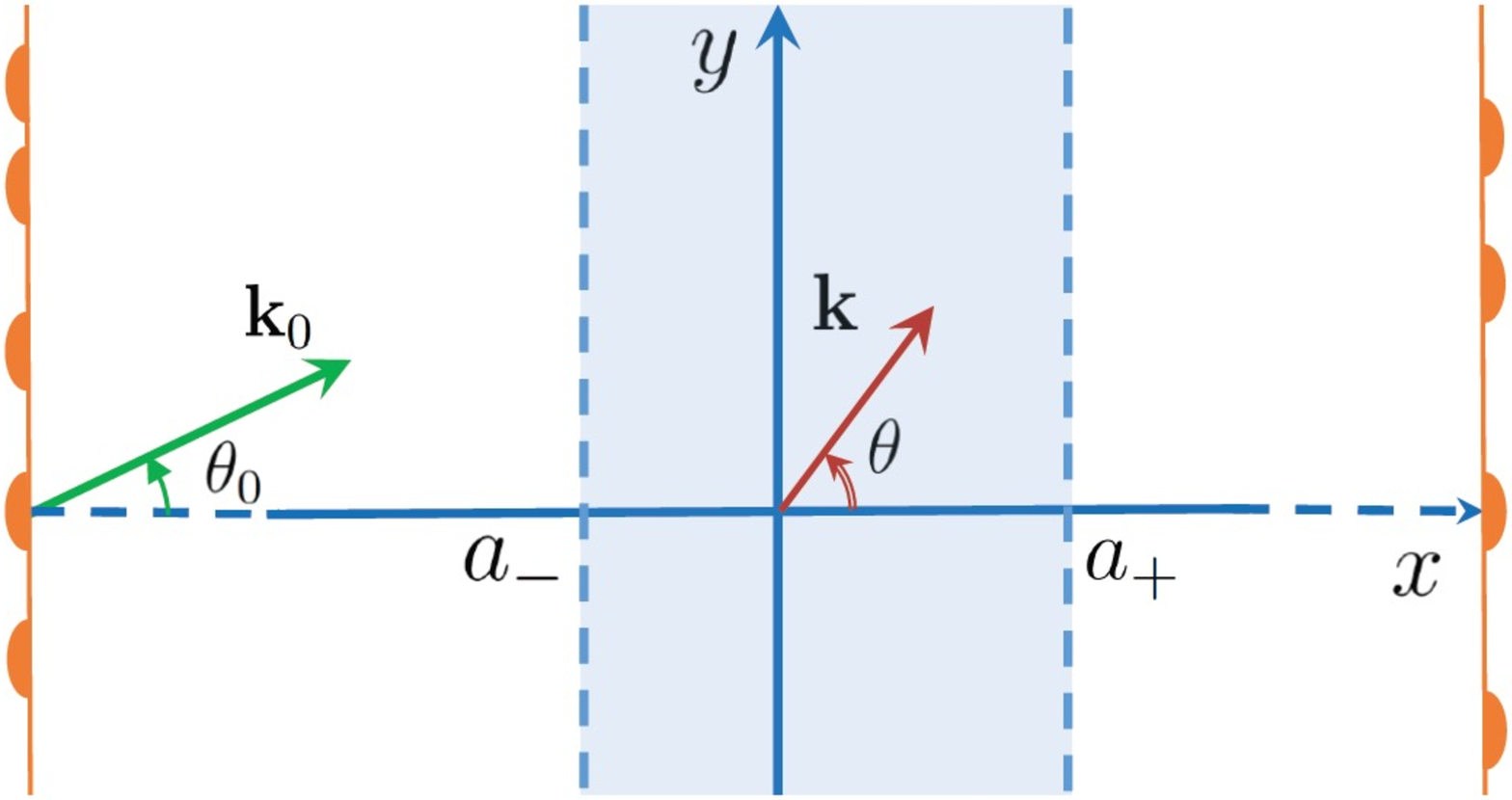}
        \caption{Schematic view of the scattering setup for a
        left-incident wave (on the left) and a right-incident
        wave (on the right). The interaction potential vanishes outside the blue region. The orange vertical lines represent the lines given by $x=\pm\infty$ on which the detectors (depicted by orange half-ellipses) are placed. $\bk_0$ and $\bk$ are respectively the incident and scattered wave vectors. For the left- and right-incident waves the incidence angle $\theta_0$ takes values in $(-\frac{\pi}{2},\frac{\pi}{2})$ and $(\frac{\pi}{2},\frac{3\pi}{2})$, respectively.}
        \label{fig1}
        \end{center}
        \end{figure}

In a scattering experiment, the source of the incident wave and the detectors measuring the scattered wave are located at spatial infinities. We choose our coordinate system in such a way that they lie along the lines $x=\pm\infty$. As we illustrate in Fig.~\ref{fig1}, the detectors may be put on both of these lines, but the source of the incident wave is either at $x=-\infty$ or $x=+\infty$.\footnote{In Fig.~\ref{fig1} we consider an interaction potential vanishing outside the region bounded by the lines given by $x=a_\pm$ for some $a_\pm\in\R$. This restriction has been enforced for future use and can be lifted for a general short-range potential by letting $a_\pm$ tend to $\pm\infty$. In particular, our choice of coordinates does not restrict the nature of the interaction potential.} These correspond to left-incident and right-incident waves whose incidence angles $\theta_0$ respectively range over the intervals $(-\frac{\pi}{2},\frac{\pi}{2})$ and  $(\frac{\pi}{2},\frac{3\pi}{2})$. We denote the corresponding scattering amplitudes by $\ff^l(\theta)$ and  $\ff^r(\theta)$, respectively. 

Let $\sF$ be the space of complex-valued functions of a real variable,  $\cF$ and $\cF^{-1}$ denote the Fourier transformation and its inverse, i.e., for all $y,p\in\R$,
	\begin{align}
	&\big(\cF f\big)(p):=\int_{-\infty}^\infty dy\,e^{-ip y} f(y),
	&&\big(\cF^{-1}f\big)(y):=\frac{1}{2\pi}\int_{-\infty}^\infty dp\,e^{ip y} f(p),
	\label{Fourier}
	\end{align}
$f\in\sF$ is a test function, and a tilde over a function of $y$ stands for its Fourier transform, so that $\tilde f(p):=\big(\cF f\big)(p)$. Performing the Fourier transform of both sides of (\ref{sch-eq}) with respect to $y$, we find
	\be
	-\tilde\psi''(x,p)+v(x,i\partial_p)\tilde\psi(x,p)
	=\varpi(p)^2\tilde\psi(x,p),~~~~~(x,p)\in\R^2,
	\label{sch-eq-FT}
	\ee
where a prime stands for differentiation with respect to $x$, $\tilde \psi(x,p)$ marks the Fourier transform of $\psi(x,y)$ with respect to $y$ evaluated at $p$, i.e., $\tilde \psi(x,p):=\big(\cF\psi(x,\cdot)\big)(p)$,
	\bea
	v(x,i\partial_p)f(p)&:=&
	\frac{1}{2\pi}\int_{-\infty}^\infty\!\!dq~\tilde v(x,p-q)f(q),
	\label{hat-v1}
	\eea
and 
	\bea
	\varpi(p)&:=&\left\{\begin{array}{ccc}
	\sqrt{k^2-p^2} & \for & |p|< k,\\
	i\sqrt{p^2-k^2} & \for & |p|\geq k.\end{array}\right.
	\label{varpi}
	\eea
	
Consider potentials $v$ that vanish outside a region bounded by a pair of lines parallel to the $y$-axis, i.e., there is an interval $[a_-,a_+]$ of real numbers such that $v(x,y)=0$ for $x\notin[a_-,a_+]$, as depicted in Fig.~\ref{fig1}. Then, $\tilde v(x,p)=0$ for $x\notin[a_-,a_+]$, and (\ref{sch-eq-FT}) gives 
	\[[\partial_x^2+\varpi(p)^2]\tilde\psi(x,p)=0~~\for~~x\notin[a_-,a_+].\] 
Solving this equation and performing the inverse Fourier transform with respect to $p$, we can write $\psi$ in the form
	\be
	\psi=\psi_{\rm os}+\psi_{\rm ev},
	\label{decompose}
	\ee
where $\psi_{\rm os},\psi_{\rm ev}:\R^2\to\C$ are a pair of functions satisfying,
	\bea
	\psi_{\rm os}(x,y)&=&
	\left\{\begin{array}{ccc}
	\displaystyle\int_{-k}^k \frac{dp}{4\pi^2\varpi(p)}[
	A_-(p) e^{i\varpi(p)x}+B_-(p) e^{-i\varpi(p)x}]e^{ip y} &\for& x\leq a_-,\\[12pt]
	\displaystyle\int_{-k}^k \frac{dp}{4\pi^2\varpi(p)}[
	A_+(p) e^{i\varpi(p)x}+B_+(p) e^{-i\varpi(p)x}]e^{ip y} &\for& x\geq a_+,
	\end{array}\right.
	\label{psi-o1}\\[6pt]
	\psi_{\rm ev}(x,y)&=&
	\left\{\begin{array}{ccc}
	\displaystyle\int_{|p|\geq k} \frac{dp}{4\pi^2\varpi(p)}
	C_-(p) e^{|\varpi(p)|x} e^{ipy}&\for& x\leq a_-,\\[12pt]
	\displaystyle\int_{|p|\geq k} \frac{dp}{4\pi^2\varpi(p)}
	C_+(p) e^{-|\varpi(p)|x} e^{ipy} &\for& x\geq a_+,
	\end{array}\right.
	\label{psi-e1}
	\eea
and $A_\pm,B_\pm,C_\pm:\R\to\C$ are functions such that
	\begin{align}
	&A_\pm(p)=B_\pm(p)=0~~\for~~|p|\geq k,
	&&C_\pm(p)=0~~\for~~|p|<k.
	\label{ABC-bound}
	\end{align}
In particular, $A_\pm$ and $B_\pm$ belong to
	\[\sF_k:=\left\{\phi\in\sF~\big|~\phi(p)=0~~\for~~x\notin(-k,k)\right\}.\]

According to (\ref{decompose}) -- (\ref{psi-e1}),
	\be
	\psi(x,y)\to
	\int_{-k}^k \frac{dp}{4\pi^2\varpi(p)}\Big[
	A_\pm(p) e^{i\varpi(p)x}+B_\pm(p) e^{-i\varpi(p)x}\Big]e^{ip y}~~~\for~~~
	x\to\pm\infty.
	\label{asym}
	\ee
This relation is generally valid for bounded solutions of the Schr\"odinger equation~(\ref{sch-eq}) provided that $v$ is a short-range potential \cite{yafaev}. It shows that $A_\pm$ and $B_\pm$ determine the asymptotic form of the solutions $\psi$ of the Schr\"odinger equation~(\ref{sch-eq}). This in turn suggests that they store the information about the scattering features of the potential. To see this, we use $A_\pm^{l/r}$ and $B_\pm^{l/r}$ to denote the coefficient functions $A_\pm$ and $B_\pm$ for the left/right incident waves. Then, as we show in Ref.~\cite{pra-2021}\footnote{Ref.~\cite{pra-2021} uses $\breve A_\pm^{l/r}$ and $\breve B_\pm^{l/r}$ for what we call $A^{l/r}_\pm$ and $B^{l/r}_\pm$, respectively.}, if we use $p_0$ to label the $y$-component of the incident wave vector $\bk_0$, so that $p_0:=k\sin\theta_0$, and notice that the $y$-component of the scattered wave vector $\bk:=k\bfr/r$ is given by $p=k\sin\theta$, we have
	\begin{align} 
	&A^l_-(p)=B^r_+(p)=2\pi\varpi(p_0)\delta(p-p_0)=2\pi\delta(\theta-\theta_0), 
	\quad\quad\quad\quad B^l_+(p)=A_-^r(p)=0,
	\label{z81}\\[6pt]
	&\ff^l(\theta)=-\frac{i}{\sqrt{2\pi}}\times \left\{
	\begin{array}{ccc}
	A^l_+(k \sin\theta)-2\pi\delta(\theta-\theta_0) &\for & \theta\in(-\frac{\pi}{2},\frac{\pi}{2}),\\[6pt]
	B^l_-(k \sin\theta) &\for & \theta\in(\frac{\pi}{2},\frac{3\pi}{2}),\end{array}\right.
	\label{z82}\\[6pt]
	&\ff^r(\theta)=-\frac{i}{\sqrt{2\pi}}\times \left\{
	\begin{array}{ccc}
	A^r_+(k \sin\theta) &\for & \theta\in(-\frac{\pi}{2},\frac{\pi}{2}),\\[6pt]
	B^r_-(k \sin\theta)-2\pi\delta(\theta-\theta_0) &\for & \theta\in(\frac{\pi}{2},\frac{3\pi}{2}).\end{array}\right.
	\label{z83}
	\end{align} 
	
In analogy with one dimension \cite{muga-review,bookchapter}, we can identify the scattering operator, also known as the S-matrix, with a $2\times 2$ matrix satisfying
	\be
	\widehat\bS\left[\begin{array}{c}
	A_-\\
	B_+\end{array}\right]=
	\left[\begin{array}{c}
	A_+\\
	B_-\end{array}\right].
	\ee 
Notice however that unlike its one-dimensional counterpart, the S-matrix $\widehat\bS$ is not a numerical matrix; it is a $2\times 2$ matrix whose entries $\widehat S_{ij}$ are linear operators acting in $\sF_k$. This makes $\widehat\bS$ a linear operator acting in the space of two-component functions,
	\[\sF_k^{2\times 1}:=
	\left\{\left.\left[\begin{array}{c}
	\phi_+\\
	\phi_-\end{array}\right]\right|\phi_\pm\in\sF_k\right\}.\]
In view of (\ref{z81}) -- (\ref{z83}), we can express the scattering amplitudes $\ff^{l/r}$ in terms of $\widehat\bS$, \cite{p167}.

The evanescent waves $\psi_{\rm ev}(x,y)$ decay as $x\to\pm\infty$, and as a result their Fourier coefficients $C_\pm(p)$ do not enter the asymptotic expression (\ref{asym}) for $\psi$ and the scattering amplitudes $\ff^{l/r}$. This does not however mean that they do not affect the outcome of the scattering calculations. This is because $C_\pm$ contribute to the value of the wave function $\psi(x,y)$ on the lines $x=a_\pm$. This makes them influence the way $A_\pm^{l/r}$ and $B_\pm^{l/r}$ relate to one another. Therefore, they do contribute to the S-matrix and the scattering amplitudes. Ignoring the contribution of $C_\pm$ to the scattering amplitudes would in general lead to errors which, depending on the behavior of the potential, may or may not be negligible. We wish to study the (propagating-wave) approximation scheme in which these errors are ignored. The outcome of this approximation turns out to be equivalent to the solution of the scattering problem for a certain nonlocal potential that does not couple to evanescent waves. As we show below, the propagating-wave approximation has certain appealing properties. For example, it produces reliable results for weak potentials and high energy incident waves. Therefore, its domain of applicability overlaps those of the first Born approximation and the semiclassical approximation. An important advantage of the propagating-wave approximation is that it is capable of providing an exact solution of the scattering problem for certain highly nontrivial complex potentials with possible optical realizations.

\section{Coupling of potentials to evanescent waves}
\label{S2}

According to (\ref{psi-o1}) and (\ref{psi-e1}), for $x\notin[a_-,a_+]$, $\psi_{\rm os}(x,y)$ is a superposition of Fourier modes $e^{ipy}$ with $|p|< k$, while $\psi_{\rm ev}(x,y)$ is a superposition of Fourier modes $e^{ipy}$
with $|p|\geq k$. To identify $\psi_{\rm os}$ and $\psi_{\rm ev}$ as functions defined in $\R^2$, we demand that $\psi_{\rm os}(x,y)$ and $\psi_{\rm ev}(x,y)$ possess the same property for $|x|\leq a$. It proves useful to use Dirac's bra-ket notation for this purpose.

For each $x\in\R$, let $|\psi(x)\kt:\R\to\C$ denote the function that assigns $\psi(x,y)$ to each $y\in\R$ according to $\br y|\psi(x)\kt:=\psi(x,y)$. Then we can express (\ref{sch-eq}) as
	\be
	[-\partial_x^2+\widehat p^{\,2}+v(x,\widehat y)]|\psi(x)\kt=k^2|\psi(x)\kt,~~~x\in\R,
	\label{sch-eq2}
	\ee
where $\widehat y$ and $\widehat p$ are the standard position and momentum operators,
	\begin{align}
	&\br y|\widehat y|\phi\kt=y\br y|\phi\kt, 
	&&\br y |\widehat p|\phi\kt:=-i\partial_y\br y|\phi\kt.\nn
	\end{align}	
As an operator acting in $L^2(\R)$, $\widehat p$ is self-adjoint and its spectrum coincides with $\R$. We use $|p\kt$ to denote its generalized eigenfunctions, where $p\in\R$, so that $\br y|p\kt=(2\pi)^{-1/2}e^{ipy}$. We denote the orthogonal projection operators associated with $\widehat p$ by $|p\kt\br p|$, introduce the projection operator, 
	\be
	\widehat\Pi_k:=\int_{-k}^k dp\:|p\kt\br p|,
	\label{Pi-k-def}
	\ee 
and use $\widehat 1$ and $\widehat 0$ to label the identity and zero operator acting in $L^2(\R)$, respectively.
	
Recalling that for each $|\phi\kt\in L^2(\R)$, $\tilde\phi(p):=\int_{-\infty}^\infty e^{-ipy}\br y|\phi\kt$ is the Fourier transform of $\br y|\phi\kt$, we can use (\ref{Pi-k-def}) to infer that
	\be
	\br y|\widehat\Pi_k|\phi\kt=\frac{1}{2\pi}\int_{-k}^kdp~e^{ipy}\tilde\phi(p).
	\label{Pi-k}
	\ee 
In view of the completeness of $|p\kt\br p|$, we also have
	\be
	\br y|(\widehat 1-\widehat\Pi_k)|\phi\kt=\frac{1}{2\pi}\int_{|p|\geq k}dp~e^{ipy}\tilde\phi(p).
	%:=\frac{1}{2\pi}\int_{-\infty}^{-k}dp~e^{ipy}\tilde\phi(p)+
	%\frac{1}{2\pi}\int_{k}^\infty dp~e^{ipy}\tilde\phi(p),
	\label{1-Pi-k}
	\ee
Furthermore, because $\widehat\Pi_k^2=\widehat\Pi_k$, 
	\begin{align}
	&\widehat\Pi_k(\widehat 1-\widehat\Pi_k)=(\widehat 1-\widehat\Pi_k)\widehat\Pi_k=\widehat 0,
	&&(\widehat 1-\widehat\Pi_k)^2=\widehat 1-\widehat\Pi_k.
	\end{align} 
	
Next, we use the projection operator $\widehat\Pi_k$ to identify the oscillating and evanescent parts, $\psi_{\rm os}$ and $\psi_{\rm ev}$, of the solutions of the Schr\"odinger equation~(\ref{sch-eq2}) as follows.
	\begin{align}
	&|\psi_{\rm os}(x)\kt:=\widehat\Pi_k|\psi(x)\kt, &&|\psi_{\rm ev}(x)\kt:=(\widehat 1-\widehat\Pi_k)|\psi(x)\kt.
	\label{z1}
	\end{align}
Clearly,
	\begin{align}
	&|\psi(x)\kt=|\psi_{\rm os}(x)\kt+|\psi_{\rm ev}(x)\kt.
	\label{z3}
	\end{align}
We also introduce,
	\begin{align}
	&\widehat V_{k}(x):=\widehat\Pi_k v(x,\widehat y)\widehat\Pi_k, &&
	\widehat W_{k}(x):=(\widehat 1-\widehat\Pi_k)\, v(x,\widehat y) \,(\widehat 1-\widehat\Pi_k),
	\label{z2}\\
	&\widehat V_-(x):=\widehat\Pi_k\,v(x,\widehat y)\,(\widehat 1-\widehat\Pi_k), &&
	\widehat V_+(x):=(\widehat 1-\widehat\Pi_k)\, v(x,\widehat y) \,\widehat\Pi_k.
	\label{z2b}
	\end{align}
Applying $\widehat\Pi_k$ and $\widehat 1-\widehat\Pi_k$ to both sides of (\ref{sch-eq2}) and using 
(\ref{z1}) and (\ref{z3}), we find
	\bea
	&&\left[-\partial_x^2+\widehat p^2+\widehat V_{k}(x)\right]|\psi_{\rm os}(x)\kt+\widehat V_-(x)|\psi_{\rm ev}(x)\kt=
	k^2|\psi_{\rm os}(x)\kt,
	\label{z5}\\
	%&&\left[-\partial_x^2+\widehat \sW_{k}(x)\right]|\psi_{\rm ev}(x)\kt+\widehat V_+(x)|\psi_{\rm os}(x)\kt=0,\\
	&&\left[-\partial_x^2+\widehat p^2+\widehat W_{k}(x)\right]|\psi_{\rm ev}(x)\kt+\widehat V_+(x)|\psi_{\rm os}(x)\kt=
	k^2|\psi_{\rm ev}(x)\kt.
	\label{z6}
	\eea
	
As seen from (\ref{z5}) the potential couples to the evanescent part of the wave through the operator $\widehat V_-(x)$. With the aid of (\ref{Pi-k}), (\ref{1-Pi-k}), and (\ref{z2b}), we can express the action of $\widehat V_-(x)$ on a test function $|\phi\kt$ in the form,
	\bea
	\widehat V_-(x)|\phi\kt&=&\int_{-k}^k dp\int_{-\infty}^{-k}dq\: \tilde v(x,p-q)\br q|\phi\kt\:|p\kt+
	\int_{-k}^k dp\int_{k}^{\infty}dq\: \tilde v(x,p-q)\br q|\phi\kt\:|p\kt.
	\label{v-minus}	\eea
Suppose that this quantity vanishes for all $\phi$. If $\phi(q)=0$ for $q\geq k$, the second term on the right-hand side of (\ref{v-minus}) is zero. This shows that for all such $\phi$, the first integral on the right-hand side of (\ref{v-minus}) must also vanish. This happens only if $\tilde v(x,p-q)=0$ for $|p|\leq k$ and $q\leq -k$.\footnote{Here we make use of the fact that because $\tilde v(x,p')$ is the Fourier transform of $v(x,y)$ with respect to $y$, it is a continuous function of $p'$.} It is easy to see that this condition is equivalent to $\tilde v(x,p')=0$ for $p'\geq 0$. Similarly, considering arbitrary test functions $\phi$ such that $\phi(q)=0$ for $q\leq k$, we can use $\widehat V_-(x)|\phi\kt=0$ to conclude that $\tilde v(x,p')=0$ for $p'\leq 0$. This argument shows that the term $\widehat V_-(x)|\psi_{\rm ev}(x)\kt$ on the left-hand side of (\ref{z5}) vanishes and the potential does not couple to the evanescent part of the wave provided that $\tilde v(x,p')=0$ for all $p'\in\R$. But this implies that $v(x,y)=0$ for all $y\in\R$, i.e., the potential vanishes. Therefore, {\em as far as the solution of the Schr\"odinger equation~(\ref{sch-eq}) is concerned, one can never neglect the coupling of a (nonzero) potential to the evanescent part of the wave}. This no-go argument does not however imply that neglecting $\widehat V_-(x)|\psi_{\rm ev}(x)\kt$ will always introduce errors in the solution of the scattering problem for the potential. This is simply because the scattering amplitudes $\ff^{l/r}$ are only sensitive to the asymptotic form of the scattering solutions of the Schr\"odinger equation, and it is in principle possible that the contribution of the term $\widehat V_-(x)|\psi_{\rm ev}(x)\kt$ to these solutions become negligible or disappear altogether as $x\to\pm\infty$.

In the propagating-wave approximation, where $\widehat V_-(x)|\psi_{\rm ev}(x)\kt$ is neglected, (\ref{z5}) reduces to
	\be
	\left[-\partial_x^2+\widehat p^2+\widehat V_{k}(x)\right]|\psi_{\rm os}(x)\kt=
	k^2|\psi_{\rm os}(x)\kt.
	\label{sch-eq-nonlocal-1}
	\ee
This is equivalent to the Schr\"odineger equation,
	\be
	\big(\widehat\bp^2+\widehat\sV_k\,\big)|\psi\kt=k^2|\psi\kt,
	\label{sch-eq-nonlocal}
	\ee
for the energy-dependent nonlocal potential, 
	\be
	\widehat\sV_k:=\widehat\Pi_k\, v(\widehat x,\widehat y)\,\widehat\Pi_k,
	\label{nonlocal-pot}
	\ee
because
	\begin{align}
	\br x,y|\widehat\sV_k|\psi\kt&:=
	\br y|\widehat\Pi_k v(x,\widehat y)\widehat\Pi_k|\psi(x)\kt=\br y|\widehat V_{k}(x)|\psi(x)\kt
	\label{nonlocal-V}\\
	&=\frac{1}{4\pi^2}\int_{-k}^k dp\int_{-k}^k dq\: e^{iyp}\,\tilde v(x,p-q)\tilde\psi(x,q)
	\label{nonlocal-V-2nd}.
	\end{align}
As $k\to\infty$, the right-hand side of (\ref{nonlocal-V-2nd}) tends to $v(x,y)\br x,y|\psi\kt$, $\widehat\sV_k\to v(\widehat x,\widehat y)$, and (\ref{sch-eq-nonlocal}) coincide with the original Schr\"odinger equation (\ref{sch-eq}). Therefore, {\em the propagating-wave approximation is valid at high energies}. In this respect it is similar to the semiclassical approximation.

Next, consider a pair of short-range potentials, $v_1$ and $v_2$, and let $\widehat \sV_{1,k}:=\widehat\Pi_k v_1(\widehat x,\widehat y)\widehat\Pi_k$ and 
$\widehat \sV_{2,k}:=\widehat\Pi_k v_2(\widehat x,\widehat y)\widehat\Pi_k$. Suppose that
	\be
	\tilde v_1(x,p)=\tilde v_2(x,p)~~~\for~~~|p|<2k.
	\label{equivalent}
	\ee
Then, in view of (\ref{nonlocal-V-2nd}), $\widehat \sV_{1,k}=\widehat \sV_{2,k}$. This implies that the application of the propagating-wave approximation for these potentials yields identical results. For example, let 
	\begin{align}
	&v_1(x,y)=g(x)\delta(y),
	&&v_2(x,y)=\frac{g(x)\sin(\fK y)}{\pi y},\nn
	\end{align} 
where $g:\R\to\C$ is a function such that $xg(x)\to 0$ as $x\to\pm\infty$, and $\fK\in\R^+$. Because
	\begin{align}
	&\tilde v_1(x,p)=g(x),
	&&\tilde v_2(x,p)=\left\{\begin{array}{ccc}
	g(x) &\for & |p|<\fK,\\
	0 & \for & |p|>\fK,\end{array}\right.\nn
	\end{align} 
$v_1$ and $v_2$ satisfy (\ref{equivalent}) for $\fK\geq 2k$. Therefore, the propagating-wave approximation does not distinguish between their scattering properties for wavenumbers $k\leq \fK/2$. This observation becomes particularly useful, for the case where $g(x)=\fz\delta(x)$ and $\fz$ is a real or complex coupling constant, i.e., when $v_1$ is a delta-function potential in two dimensions, because for this potential the propagating-wave approximation turns out to give the exact expression for the scattering amplitude. This is actually not an exclusive feature of the delta-function potential; there is a large class of potentials for which the propagating-wave approximation is exact. In the remainder of this article we employ the dynamical formulation of stationary scattering (DFSS) of Ref.~\cite{pra-2021} to identify these potentials and arrive at a better understanding of the propagating-wave approximation.

\section{Dynamical formulation of stationary scattering}
\label{S3}

In one dimension, there is an alternative to the S-matrix, called the transfer matrix, which also stores the information about the scattering properties of the potential \cite{griffiths,sanchez,tjp-2020}. In addition, it enjoys a useful composition property which allows for the calculation of the scattering properties of a short-range potential $v$ using the scattering properties of a finite number of its truncations $v_i$ that add up to $v$ and have smaller non-overlapping supports \cite{tjp-2020}.\footnote{For a discussion of the generalization of the transfer matrix for long-range potentials in one dimensions, see \cite{jpa-2020b}.}  This feature of the transfer matrix is the main reason for its wide range of applications \cite{jones-1941,abeles,thompson,teitler-1970,berreman-1972,yeh,abrahams-1980,ardos-1982,pendry-1982,levesque,hosten,sheng-1996,schubert-1996,wang-2001,wortmann-2002,katsidis-2002,yeh-book,Hao-2008,li-2009,zhan-2013}. The practical advantages of using the transfer matrix in one dimensions has motivated the development of its multichannel \cite{pereyray-1998a,pereyray-2002,pereyray-2005,Shukla-2005,anzaldo-meneses-2007}
and higher-dimensional generalizations
\cite{pendry-1984,pendry-1990a, pendry-1990b,pendry-1994,mclean,ward-1996,pendry-1996}. The latter involve a discretization of either the configuration or momentum space variables along the normal directions to the principle scattering/propagation axis and yield large numerical transfer matrices whose treatment requires appropriate numerical schemes.

Ref.~\cite{pra-2021} pursues a different route to obtain a higher-dimensional notion of transfer matrix which does not require any discretization (and a corresponding approximation) scheme.\footnote{See also Ref.~\cite{pra-2016}.} Its point of departure is a natural higher-dimensional extension of the definition of the transfer matrix in one dimension. Specifically, in two dimensions, it identifies the transfer matrix with a $2\times 2$ matrix $\widehat\bM$ with operator entries $\widehat M_{ij}:\sF_k\to\sF_k$ such that
	\be
	\widehat\bM\left[\begin{array}{c}
	A_-\\
	B_-\end{array}\right]=
	\left[\begin{array}{c}
	A_+\\
	B_+\end{array}\right].
	\label{M-def}
	\ee
If we insert $A^{l/r}_\pm$ and $B^{l/r}_\pm$ respectively for $A_\pm$ and $B_\pm$ in this relation and make use of (\ref{z81}), we arrive at 
	\begin{align}
	&\widehat M_{22}\,B^l_-=-2\pi\widehat M_{21}\:\delta_{p_0},
	&& \widehat M_{22}\,B^r_-=2\pi \delta_{p_0}
	\label{eq1a}\\
	&A^l_+= 2\pi \widehat M_{11}\:\delta_{p_0}+\widehat M_{12}B_-^l,
	&& A^r_+=\widehat M_{12}B^r_-,
	\label{eq1b}
	\end{align}
where $\delta_{p_0}$ stands for the Dirac delta function centered at $p_0$, i.e., $\delta_{p_0}(p):=\delta(p-p_0)$. Notice that being the $y$-component of the incident wave vector $\bk_0$, $p_0$ satisfies $p_0=k\sin\theta_0$. In particular, $|p_0|<k$. 

Eqs.~(\ref{eq1a}) and (\ref{eq1b}) provide a method for calculating the scattering amplitude of the potential; solving (\ref{eq1a}) for $B^{l/r}_-$, using the result in (\ref{eq1b}) to determine $A^{l/r}_+$, and substituting $A^{l/r}_+$ and $B^{l/r}_-$ in (\ref{z82}) and (\ref{z83}), we can calculate $\ff^{l/r}$, \cite{pra-2021}.

A remarkable property of the transfer matrix $\widehat\bM$ is that, similarly to its well-known one-dimensional counterpart \cite{ap-2014,pra-2014a}, it can be expressed in terms of the evolution operator for an effective non-unitary quantum system \cite{pra-2021}. This requires the introduction of an auxiliary transfer matrix $\widehat\bfM$ which satisfies,
	\be
	\widehat\bfM\left[\begin{array}{c}
	A_-\\
	\sB_-\end{array}\right]=
	\left[\begin{array}{c}
	\sA_+\\
	B_+\end{array}\right],
	\label{fM-def}
	\ee
where
	\begin{align}
	&\sB_-(p)=B_-(p)+C_-(p)=\left\{\begin{array}{ccc}
	B_-(p)&\for&|p|<k,\\
	C_-(p)&\for&|p|\geq k,\end{array}\right.
	\label{sB-}\\
	&\sA_+(p)=A_+(p)+C_+(p)=\left\{\begin{array}{ccc}
	A_+(p)&\for&|p|<k,\\
	C_+(p)&\for&|p|\geq k.\end{array}\right.
	\label{sA+}
	\end{align}
According to (\ref{fM-def}), we should view $\widehat\bfM$ as a linear operator acting in the space of two-component functions,
	\[\sF^{2\times 1}:=
	\left\{\left.\left[\begin{array}{c}
	\xi_+\\
	\xi_-\end{array}\right]\right|\xi_\pm\in\sF\right\}.\]

The auxiliary transfer matrix has two important properties \cite{pra-2021}:
	\begin{enumerate}
	\item It admits an expression in terms of the evolution operator $\widehat\bcU(x,x_0)$ for the Hamiltonian operator,		
	\be
   	\widehat\bcH(x):=\frac{1}{2}\,e^{-i\widehat\varpi x\bsigma_3}\,
         v(x,\widehat y)\,\widehat\varpi ^{-1}\bcK\, e^{i\widehat\varpi x\bsigma_3},
         \label{bcH-def}
        \ee
where $x$ plays the role of time, $\widehat\varpi:=\varpi (\widehat p)$, $\widehat p$ is the $y$-component of the standard momentum operator, 
	\be
	\bcK:=\left[\begin{array}{cc}
    1 & 1 \\
    -1 & -1\end{array}\right]=\bsigma_3+i\bsigma_2,
    \label{bcK-def}
    \ee
and $\bsigma_j$ denote the Pauli matrices;
	\begin{align}
	&\bsigma_1:=\left[\begin{array}{cc}
	0 & 1\\
	1 & 0\end{array}\right],
	&& \bsigma_2:=\left[\begin{array}{cc}
	0 & -i\\
	i & 0\end{array}\right],
	&&\bsigma_3:=\left[\begin{array}{cc}
	1 & 0\\
	0 & -1\end{array}\right].
	\label{Pauli}
	\end{align}
The evolution operator $\widehat\bcU(x,x_0)$ for the Hamiltonian~(\ref{bcH-def}) gives the auxiliary transfer matrix according to $\widehat\bfM=\widehat\bcU(a_+,a_-)$. In particular, employing the Dyson series expansion of $\bcU(x,x_0)$ and noting that $\widehat{\bcH}(x)=\bzero$ for $x\notin[a_-,a_+]$, we have	
	\be
	\widehat\bfM=\widehat\bI+\sum_{n=1}^\infty (-i)^n
         \int_{-\infty}^{\infty} \!\!dx_n\int_{-\infty}^{x_n} \!\!dx_{n-1}
         \cdots\int_{-\infty}^{x_2} \!\!dx_1\,
         \widehat{\bcH}(x_n)\widehat{\bcH}(x_{n-1})\cdots\widehat{\bcH}(x_1).
         \label{bcM-Dyson}
	\ee
%Ref.~\cite{pra-2021} uses this relation to establish the composition property of $\widehat\bfM$ which is reminiscent of the well-known composition property of the transfer matrix of scattering theory in one dimension \cite{tjp-2020}. 

\item Let $\widehat\Pi_k:\sF\to\sF$ be the following extension of the projection operator (\ref{Pi-k-def}) to $\sF$,
	\be
	\big(\widehat\Pi_k\xi\big)(p):=\left\{\begin{array}{ccc}
	\xi(p)&\for&|p|<k,\\
	0 &\for& |p|\geq k,\end{array}\right.
	\label{project1}
	\ee
and $\widehat\bPi_k:\sF^{2\times 1}\to\sF^{2\times 1}$ be the projection operator defined by
	\begin{align}
	&\widehat\bPi_k\left[\begin{array}{c}\xi_+\\ \xi_-\end{array}\right]:=
    	\left[\begin{array}{c}\widehat\Pi_k\xi_+\\ \widehat\Pi_k\xi_-\end{array}\right].
	\label{project2}
	\end{align}
Then, we can express the transfer matrix $\widehat\bM$ in terms of $\widehat\bfM$ and $\widehat\bPi_k$ according to
	\be
    	\widehat\bM=\widehat\bPi_k\,\widehat\bfM\,\widehat\bPi_k.
    	\label{M-M}
    	\ee
	\end{enumerate}
	
The presence of the operator $v(x,\widehat y)$ on the right-hand side of (\ref{bcH-def}) shows that if we scale the potential as 
	\be
	v(x,y)\to\alpha\,v(x,y),
	\label{scale}
	\ee 
for some $\alpha\in\R^+$, then the effective Hamiltonian also scales by a factor of $\alpha$. This in turn allows us to view the Dyson series (\ref{bcM-Dyson}) as a power series in the strength of the potential. Affecting the scaling transformation (\ref{scale}) in (\ref{bcM-Dyson}), we find a series in powers of $\alpha$ which we can identify as a perturbation series. Suppose that we neglect all but the first $N+1$ terms of the series in (\ref{bcM-Dyson}) and substitute the result in (\ref{M-M}) to determine an approximate expression for the transfer matrix $\widehat\bM$. If we use this expression to solve (\ref{eq1a}) and (\ref{eq1b}) for $B_-^{l/r}$ and $A_+^{l/r}$ and insert the outcome in (\ref{z82}) and (\ref{z83}), we obtain approximate formulas for the scattering amplitudes $\ff^{l/r}$. If we expand these formulas in powers of $\alpha$, neglect the terms of order $\alpha^{N+1}$ and higher, and finally set $\alpha=1$, we recover the result of the $N$-th Born approximation. In particular, the first Born approximation corresponds to setting $N=1$. In this case, (\ref{bcM-Dyson})  and (\ref{M-M}) give
	\bea
	\widehat\bM&\approx&\widehat\bPi_k-i\int_{-\infty}^\infty dx\:\widehat\bcH_k(x),
	\label{M-N=1}
	\eea
where
	\be
	\widehat\bcH_k(x):=\widehat\bPi_k\widehat\bcH(x)\widehat\bPi_k.
	\label{bcH-k}
	\ee
In view of (\ref{z2}), (\ref{bcH-def}), (\ref{project1}), (\ref{project2}), (\ref{bcH-k}), and the fact that $[\widehat p,\widehat\Pi_k]=[\widehat\varpi,\widehat\Pi_k]=\widehat 0$,
	\bea
	\widehat\bcH_k(x)&=&
	\frac{1}{2}\,e^{-i\widehat\varpi x\bsigma_3}\,
         \widehat\Pi_k\,v(x,\widehat y)\,\widehat\Pi_k
         \widehat\varpi ^{-1}\,\bcK\, e^{i\widehat\varpi x\bsigma_3}\nn\\
         &=&\frac{1}{2}\,e^{-i\widehat\varpi x\bsigma_3}\,
         \widehat V_{k}(x)\,\widehat\varpi ^{-1}\bcK\, e^{i\widehat\varpi x\bsigma_3}.\label{eq52}
         \eea
Substituting this relation in the right-hand side of (\ref{M-N=1}) and recalling (\ref{nonlocal-V}), we can identify the resulting approximate expression for $\widehat\bM$ as the one we would obtain, if we let the nonlocal potential $\widehat\sV_k$ play the role of the original potential $v(\widehat x,\widehat y)$.  This argument shows that {\em the propagating-wave approximation is consistent with the first Born approximation, i.e., it is a valid approximation for weak potentials}.

Next, we examine the utility of DFSS in the study of the scattering properties of the nonlocal potentials $\widehat\sV_k$. Then, in view of (\ref{nonlocal-V}), $\widehat\bcH(x)=\widehat\bcH_k(x)$. In particular, $\widehat\bPi_k \widehat\bcH(x)\widehat\bPi_k=\widehat\bcH(x)$. This equation together with (\ref{bcM-Dyson}) and (\ref{M-M}) show that for the nonlocal potentials $\widehat\sV_k$, the (fundamental) transfer matrix $\widehat\bM$ coincides with the auxiliary transfer matrix $\widehat\bfM$. Therefore, for these potentials, 
	\be
	\widehat\bM=\widehat\bI+\sum_{n=1}^\infty (-i)^n
         \int_{-\infty}^{\infty} \!\!dx_n\int_{-\infty}^{x_n} \!\!dx_{n-1}
         \cdots\int_{-\infty}^{x_2} \!\!dx_1\,
         \widehat{\bcH}_k(x_n)\widehat{\bcH}_k(x_{n-1})\cdots\widehat{\bcH}_k(x_1).
         \label{bM-Dyson-Hk}
	\ee

\section{Exactness of the propagating-wave approximation}

In Ref~\cite{pra-2021} we show that for potentials of the form $v(x,y)=\delta(x)\fg(y)$ with $\fg\in\sF$, (\ref{M-N=1}) holds as an exact equality.\footnote{A simple example is the delta-function potential $v(x,y)=\fz\,\delta(x)\delta(y)$.} Therefore for these potentials the propagating-wave approximation provides the exact expression for the scattering amplitudes. The following result identifies an infinite class of complex potentials with this property.\\[6pt]
\noindent {\bf Theorem}: Let $v:\R^2\to\C$ be a potential such that its Fourier transform with respect to $y$ vanishes on one of the half-axes given by $\pm p\leq 0$, i.e., either
	\be
	\tilde v(x,p)=0~~~\for~~~p\leq 0,
	\label{thm-condi}
	\ee
or
	\be
	\tilde v(x,p)=0~~~\for~~~p\geq 0.
	\label{thm-condi-p}
	\ee 
Then the propagating-wave approximation is exact for $v$.\\[6pt]
To prove this theorem, first we use (\ref{bcM-Dyson}) and (\ref{M-M}) to express the fundamental transfer matrix in the form,
	\be
	\widehat\bM=\widehat\bI+\sum_{n=1}^\infty (-i)^n
         \int_{-\infty}^{\infty} \!\!dx_n\int_{-\infty}^{x_n} \!\!dx_{n-1}
         \cdots\int_{-\infty}^{x_2} \!\!dx_1\,
         \widehat\bPi_k
         \widehat{\bcH}(x_n)\widehat{\bcH}(x_{n-1})\cdots\widehat{\bcH}(x_1)\widehat\bPi_k.
         \label{bM-Dyson}
         \ee
According to (\ref{bcH-def}) and (\ref{eq52}), the propagating-wave approximation is exact, if we can replace the $\widehat{\bcH}(x_{j})$'s in this equation with $\widehat{\bcH}_k(x_{j})$.  To arrive at a more explicit description of this condition, we derive an alternative expression for the product of $\widehat{\bcH}(x_{j})$'s.

Consider the operators,
	\begin{align}
	&\widehat s(x):=\widehat\varpi^{-1}
	\sin(x\,\widehat\varpi),
	&&
	\widehat\bcL(x):=e^{-ix\widehat\varpi\bsigma_3}
	\left[\begin{array}{cc}
	0 & 0 \\
	1 & 1\end{array}\right]e^{ix\widehat\varpi\bsigma_3}=
	\left[\begin{array}{cc}
	0 & 0 \\
	e^{2ix\widehat\varpi} & 1\end{array}\right],
	\label{sm-def}
	\end{align}
which respectively act in $\sF$ and $\sF^{2\times 1}$. Then a straightforward usage of the properties of the Pauli matrices and Eqs.~(\ref{bcH-def}) and (\ref{bcK-def}) allows us to show that 
	\be
	\widehat\bcH(x_n)\widehat\bcH(x_{n-1})\cdots\widehat\bcH(x_1)=
	\frac{i^n}{2}\, e^{-ix_n\widehat\varpi\bsigma_3}
	\widehat \cH_n(x_n,x_{n-1},\cdots,x_1)\,\bcK\,
e^{ix_1\widehat\varpi\bsigma_3}\widehat\varpi^{-1}\widehat\bcL(x_1),
	\label{lemma2}
	\ee
where $\widehat\cH_1(x_1):=v(x_1,\widehat  y)$, and for $n\geq 2$,
	\begin{align}
	\widehat\cH_n(x_n,x_{n-1},\cdots,x_1):=&
	v(x_n,\widehat  y)\, \widehat s(x_n\!-\!x_{n-1})
	v(x_{n-1},\widehat  y)\,\widehat s(x_{n-1}\!-\!x_{n-2})\,
	v(x_{n-2},\widehat  y)\,\cdots \nn\\
	&\cdots\, v(x_2,\widehat  y)\,
	\widehat s(x_{2}\!-\!x_{1})\,v(x_1,\widehat  y).
	\label{scalar-H}
	\end{align}
Notice that $\widehat\cH_n(x_n,x_{n-1},\cdots,x_1)$'s act in $\sF$ and contain all the information about the potential.\footnote{The presence of $\widehat\bcL(x_1)$ on the right-hand side (\ref{lemma2}) is necessary for the correct identification of the domain of definition of $\widehat\bcH(x_n)\widehat\bcH(x_{n-1})\cdots\widehat\bcH(x_1)$, \cite{p171}. It does not however play a role in the proof of the above Theorem.} Therefore, in order to establish the exactness of the propagating-wave approximation for potentials fulfilling (\ref{thm-condi}) or (\ref{thm-condi-p}), it suffices to show that
	\begin{align}
	\widehat\Pi_k\widehat\cH_n(x_n,x_{n-1},\cdots,x_1)\widehat\Pi_k=&
	\widehat V_k(x_n)\, \widehat s(x_n\!-\!x_{n-1})
	\widehat V_k(x_{n-1})\,\widehat s(x_{n-1}\!-\!x_{n-2})\,
	\widehat V_k(x_{n-2})\,\cdots \nn\\
	&\cdots\,
	\widehat V_k(x_2)\,
	\widehat s(x_{2}\!-\!x_{1})\,\widehat V_k(x_1).
	\label{relation-thm}
	\end{align}
This holds trivially for $n=1$. We give its proof for $n\geq 2$ in the appendix for potentials satisfying (\ref{thm-condi}). This completes the proof of the exactness of propagating-wave approximation for these potentials. 

Next, suppose that $v(x,y)$ fulfills (\ref{thm-condi-p}), and let $w(x,y):=v(x,-y)$. Then $\tilde w(x,p)=\tilde v(x,-p)=0$ for $p\leq 0$. Therefore, propagating-wave approximation is exact for $w$. Because the solution of the scattering problem for $v$ can be easily mapped\footnote{If $f_v^{l/r}(\theta_0,\theta)$ and $f_w^{l/r}(\theta_0,\theta)$ respectively denote the scattering amplitudes of the potentials $v$ and $w$. Then 
	\begin{align}
	&f^l_v(\theta_0,\theta)=\left\{\begin{aligned} 
	&f^l_w(-\theta_0,-\theta) &&\!\for~\theta\in\mbox{$(-\frac{\pi}{2},\frac{\pi}{2})$},\\
	&f^l_w(-\theta_0,2\pi-\theta) &&\!\for~\theta\in\mbox{$(\frac{\pi}{2},\frac{3\pi}{2})$},
	\end{aligned} \right.
	&&f^r_v(\theta_0,\theta)=\left\{\begin{aligned} 
	&f^r_w(2\pi-\theta_0,-\theta) &&\!\for~\theta\in\mbox{$(-\frac{\pi}{2},\frac{\pi}{2})$},\\
	&f^r_w(2\pi-\theta_0,2\pi-\theta) &&\!\for~\theta\in\mbox{$(\frac{\pi}{2},\frac{3\pi}{2})$}.
	\end{aligned} \right.\nn 
	\end{align}}
to that of $w$, this implies exactness of the propagating-wave approximation for $v$.

In one dimension, short-range potentials whose Fourier transform vanishes on the negative or positive half-axis have the remarkable property of being unidirectionally invisible for all frequencies \cite{horsley-2015,longhi-2015,horsley-longhi,jiang-2017}. The above theorem reveals the exactness of propagating-wave approximation for the treatment of their two-dimensional analogs. Constructing concrete examples of the latter is quite easy; given $u,g\in\sF$, the potential defined by $v(x,y):=g(x)\int_{0}^\infty dp\:e^{ipy}u(p)$ fulfills (\ref{thm-condi}). We can also select $g$ and $u$ in such a way that $v$ is a short-range potential. For instance, suppose that $g$ has a compact support, i.e., there are $a_\pm\in\R$ such that $a_-<a_+$, $g(x)=0$ for $x\notin [a_-,a_+]$, and $u(p):= p^\ell e^{-\beta|p|}/\ell !$ where $\ell$ is a positive integer and $\beta$ is a positive real parameter. Then,
	\be
 	v(x,y)=\frac{g(x)}{(\beta-iy)^{\ell+1}},
	\ee 
which is a short-range potential. By construction it satisfies (\ref{thm-condi}) and the propagating-wave approximation provides an exact description of its scattering properties. Notice also that whenever $g$ is a real-valued even function, this potential is $\cP\cT$-symmetric.

Next, consider a potential $v(x,y)$ that does not satisfy (\ref{thm-condi}). Then the propagating-wave approximation is not exact, but it may still provide a reliable approximate description of the scattering features of the potential. For example, consider a potential of the form
	\begin{align}
	&v(x,y)=g(x)\,e^{i\alpha y}e^{-\beta^2 y^2/2},
	&&g(x):=\left\{\begin{array}{cc}
	\fz & \for~x\in (a_-,a_+),\\
	0 & {\rm otherwise},\end{array}\right.
	\label{gaussian}
	\end{align}
where $\alpha$ and $\beta$ are positive real parameters, and $\fz$ is a real or complex coupling constant. Then $\tilde v(x,p)=\sqrt{2\pi}\,g(x)\beta^{-1}e^{-(p-\alpha)^2/2\beta^2}$. For $p\leq 0$, this implies $|\tilde v(x,p)|\leq |\fz|\,\beta^{-1}e^{-\alpha^2/2\beta^2}$. Therefore, if $\alpha\gg\beta\gtrapprox |\fz|$, $\tilde v(x,p)\approx 0$ for $p\leq 0$, and the propagating-wave approximation is expected to be reliable.

\section{Practical and mathematical aspects of propagating-wave approximation}

At first glance there seems to be no major difference between practical aspects of the dynamical formulation of scattering for a given short-range potential $v$ and the corresponding nonlocal potential $\widehat\sV_k$. The calculation of the transfer matrix $\widehat\bM$ for both of these potentials amounts to summing up certain Dyson series. But there is a very important difference between the Dyson series expansion of the transfer matrices for $v$ and $\widehat\sV_k$. As we demonstrate in Ref.~\cite{p171}, this has to do with the fact that we can setup the dynamical formulation of stationary scattering for $\widehat\sV_k$ in the space $L^2(-k,k)$ of square-integrable complex-valued functions defined on the interval $(-k,k)$. This already implies that one can develop effective numerical schemes for computing the transfer matrix for $\widehat\sV_k$, because they only need to deal with functions having a finite domain, namely $(-k,k)$.

A more important advantage of dealing with the nonlocal operator $\widehat\sV_k$ is that it defines a normal Hilbert-Schmidt (in particular compact) operator acting in $L^2(-k,k)$, \cite{p171}. This implies that there is an orthonormal basis of $L^2(-k,k)$ consisting of the eigenvectors of $\widehat\sV_k$, the spectrum of $\widehat\sV_k$ consists only of eigenvalues \cite{Beauzamy}, and the nonzero eigenvalues of $\widehat\sV_k$ are finitely degenerate and have $0$ as their accumulation point. Another benefit of dealing with $\widehat\sV_k$ is that the operator $\widehat s(x)$ that enters the calculation of the product of the effective Hamiltonians $\widehat\bcH_k(x_j)$ through (\ref{lemma2}) and (\ref{scalar-H})
becomes a bounded self-adjoint operator acting in $L^2(-k,k)$. This in turn implies that the operators $\widehat\cH_n(x_n,x_{n-1},\cdots,x_1)$, which determine the Dyson series expansion of the transfer matrix for $\widehat\sV_k$, are Hilbert-Schmidt and hence compact operators acting in $L^2(-k,k)$. A basic result of the theory of compact operators is that they can be approximated by finite-range operators \cite{kato,Reed-Simon1} which in effect means that we can develop accurate approximate descriptions of these operators using finite numerical matrices \cite{ahues,guven}. This observation opens up a new research front for effective numerical treatments of the scattering problem for $\widehat\sV_k$, i.e., the application of the propagating-wave approximation.

Another implication of the desirable mathematical properties of the operators $\widehat\bcH_k(x_j)$ is the development of a mathematically rigorous dynamical formulation of stationary scattering for the nonlocal potentials $\widehat\sV_k$,  \cite{p171}. This means dealing with the notorious domain issues of the related unbounded operators and proving the existence of the transfer matrix as a densely-defined operator.%, i.e., the strong convergence of the Dyson series in (\ref{bM-Dyson-Hk}) on a dense subset of $\sH$.

\section{Concluding Remarks} 

Evanescent waves are at the heart of the major differences between wave propagation in one and higher-than-one dimensions. In this article, we have explored the consequences of ignoring the contribution of the evanescent waves to the scattering properties of a given short-range potential in two dimensions. This has led us to introduce a nonperturbative approximation scheme in which the potential is replaced with an energy-dependent nonlocal potential $\widehat\sV_k$ that does not couple to the evanescent waves. The scattering solutions of the Schr\"odinger equation for this potential have purely oscillating Fourier modes. Therefore they correspond to propagating waves. 

The requirement that $\widehat\sV_k$ is nonlocal stems from the fact that every local potential necessarily couples to the evanescent waves. This does not however mean that this coupling always affects the scattering properties of the potential; there are potentials for which the propagating-wave approximation is exact. Specially, we have shown that whenever the Fourier transform of a potential $v(x,y)$ with respect to $y$ vanishes on the negative or positive real axis, the propagating-wave approximation produces the exact expression for its scattering amplitudes. It is quite remarkable that enforcing the very same condition in one dimension ensures the unidirectional invisibility of the potential at all frequencies \cite{horsley-2015,longhi-2015,horsley-longhi}!

The propagating-wave approximation has important practical advantages, for its numerical implementations can benefit from the fact that the functions entering the propagating-wave scattering calculations have a common finite domain, namely the interval $(-k,k)$, and the operators appearing in the expression for the transfer matrix can be accurately approximated by matrices. These observations provide ample motivation for developing effective numerical schemes for performing propagating-wave approximation.  

Finally, we wish to note that the validity of the results we have reported in this article are not confined to short-range potentials. For example, they apply to potentials whose support lies between a pair of lines that are parallel to the $y$-axis, such as those considered in Ref.~\cite{p167}. We can also use the machinery of the dynamical formulation of stationary scattering in three dimensions \cite{pra-2021} to arrive at the three-dimensional extensions of these results.

\section*{Acknowledgements}
This work has been supported by the Scientific and Technological Research Council of Turkey (T\"UB\.{I}TAK) in the framework of the project 120F061 and by Turkish Academy of Sciences (T\"UBA).

%\pagebreak

\section*{Appendix: Proof of Eq.~(\ref{relation-thm}) for $n\geq 2$}

Suppose that $v$ is a potential satisfying (\ref{thm-condi}). Then to 
prove (\ref{relation-thm}) for $n\geq 2$, we proceed as follows. \begin{enumerate}
\item We use (\ref{scalar-H}) to infer that for every $m\geq 2$ and $x_1,x_2,\cdots,x_m\in\R$,
	\begin{align}
	\widehat\Pi_k\widehat\cH_m(x_m,x_{m-1},\cdots,x_1)\widehat\Pi_k=\widehat\Pi_k
	\widehat\cH_{m-1}(x_m,x_{m-1},\cdots,x_2)\widehat s(x_{2}\!-\!x_{1})\,v(x_1,\widehat  y)\widehat\Pi_k.
	\label{thm-condi4a}
	\end{align}
This relation suggests that if we can verify
	\be
	\widehat\Pi_k
	\widehat\cH_{m-1}(x_m,\cdots,x_2)\widehat s(x_{2}\!-\!x_{1})\,v(x_1,\widehat  y)\widehat\Pi_k=
	\widehat\Pi_k
	\widehat\cH_{m-1}(x_m,\cdots,x_2)
	\widehat\Pi_k
	\widehat s(x_{2}\!-\!x_{1})\,v(x_1,\widehat  y)\widehat\Pi_k,
	\label{thm-condi4}
	\ee
then by repeated use of this equation and (\ref{thm-condi4a}) we can prove (\ref{relation-thm}).
	
\item For each real number $\alpha$, we introduce $\cS_\alpha:=\left\{\phi\in\sF~|~\phi(p)=0~\for~p\leq\alpha\right\}$ and show that whenever (\ref{thm-condi}) holds, $f(\widehat p)v(x,\widehat y)$ maps $\cS_\alpha$ to $\cS_\alpha$ for every $f\in\sF$. To see this we observe that for all $\phi\in\cS_\alpha$, 
	\be
	\br p|f(\widehat p)v(x,\widehat y)|\phi\kt=f(p)v(x,i\partial_p)\phi(p)=\frac{f(p)}{2\pi}\int_{-\infty}^\infty\!\! dq\:\tilde v(x,p-q)\phi(q)=\frac{f(p)}{2\pi}\int_{\alpha}^\infty\!\! dq\:\tilde v(x,p-q)\phi(q),\nn
	\ee
where we have made use of (\ref{hat-v1}). If $p\leq\alpha$, $p-q\leq 0$, (\ref{thm-condi}) implies $\tilde v(x,p-q)=0$, and we obtain $\br p|f(\widehat p)v(x,\widehat y)|\phi\kt=0$. Therefore, $f(\widehat p)v(x,\widehat y)\phi\in\cS_\alpha$. 
 
	\item We observe that for all $\phi\in\sF$, $\widehat\Pi_k\phi
	\in\cS_{-k}$. Moreover, because $\widehat s(x)=\widehat\varpi^{-1}\sin(x\,\widehat\varpi)$ and $\widehat\varpi=\varpi(\widehat p)$, 
	$\widehat s(x)$ is a function of $\widehat p$. This implies that for potentials satisfying (\ref{thm-condi}), \linebreak $\widehat s(x_{2}\!-\!x_{1})\,v(x_1,\widehat  y)\widehat\Pi_k\phi\in\cS_{-k}$. By virtue of this relation and the fact that $(\widehat 1-\widehat\Pi_k)\xi=0$ for $\xi\in\cS_{-k}$, we have
	$(\widehat 1-\widehat\Pi_k)\widehat s(x_{2}\!-\!x_{1})\,v(x_1,\widehat  y)\widehat\Pi_k\phi=0$.
Since $\phi$ is arbitrary, this is equivalent to
	\[\widehat s(x_{2}\!-\!x_{1})\,v(x_1,\widehat  y)\widehat\Pi_k=
	\widehat\Pi_k\widehat s(x_{2}\!-\!x_{1})\,v(x_1,\widehat  y)\widehat\Pi_k.\]
This equation clearly implies (\ref{thm-condi4}). 
\end{enumerate}

\ed
\begin{thebibliography}{99}

\bibitem{ford-1959} K.~W.~Ford and J.~A.~Wheeler,
``Semiclassical description of scattering,''
Ann.\ Phys.\ (NY) {\bf 7}, 259-286 (1959).

\bibitem{berry-1972}
M.~V.~Berry and K.~E.~Mount,
``Semiclassical approximation in wave mechanics,''
Rep.\ Prog.~Phys.\ {\bf 35}, 315-397 (1972).

\bibitem{koeling-1975}
T.~Koeling and R.~A.~Malfliet,
``Semi-classical approximations to heavy ion scattering
Based on the Feynman path-integral method,''
Phys.\ Rep.~{\bf 22}, 181-213 (1975)

\bibitem{adhikari-2008}
S.~K.~Adhikari,
``Semiclassical scattering in two dimensions,''
Am.\ J.~Phys.\ {\bf 76}, 1108-1113 (2008).

\bibitem{yafaev}
D.~R.~Yafaev, {\em Mathematical Scattering Theory} (AMS, Providence, 2010).

\bibitem{pra-2021} 
F.~Loran and A.~Mostafazadeh,
``Fundamental transfer matrix and dynamical formulation of stationary scattering in two and three dimensions,'' Phys.\ Rev~A {\bf 104}, 032222 (2021).

\bibitem{muga-review}
J.~G.~Muga, J.~P.~Palao, B.~Navarro, and I.~L.~Egusquiza, 
``Complex absorbing potentials,''
Phys.\ Rep.~{\bf 395}, 357-426 (2004).

\bibitem{bookchapter}
A.~Mostafazadeh,
``Scattering theory and PT-symmetry,'' in {\em Parity-Time Symmetry and Its Applications},
edited by D. Christodoulides and J. Yang, pp 75-121
(Springer, Singapore, 2018), arXiv:1711.05450.

\bibitem{p167} 
F.~Loran and A.~Mostafazadeh,
``Exceptional points and pseudo-Hermiticity in real potential scattering,'' SciPost Phys.~{\bf 12}, 109 (2022).

\bibitem{griffiths} 
D.~J.~Griffiths and C.~A.~Steinke, 
``Waves in locally periodic media,''
Am.\ J.~Phys.~{\bf 69}, 137-154 (2001).

\bibitem{sanchez}
L.~ L.~S\'{a}nchez-Soto, J.~J.~Monz\'{o}na, A.~G.~ Barriuso, and J.~ F.~Cari\~{n}ena, ``The transfer matrix: A geometrical perspective,''
Phys.\ Rep.\ {\bf 513}, 191-227 (2012).

\bibitem{tjp-2020} 
A.~Mostafazadeh, 
``Transfer matrix in scattering theory: A survey of basic properties and recent developments,'' Turkish J.~Phys.\ {\bf 44}, 472-527 (2020).

\bibitem{jones-1941}
R.~C.~Jones, ``A new calculus for the treatment of optical systems I. Description and discussion of the Calculus,''
J.\ Opt.\ Soc.\ Am.\ {\bf 31}, 488-493 (1941).

\bibitem{abeles}
F.~Abel\`es,
``Recherches sur la propagation des ondes \'electromagn\'etiques sinuso\"idales dans les milieux stratif{\i}\'es Application aux couches minces,''
Ann.\ Phys.\ (Paris) {\bf 12}, 596-640 (1950).

\bibitem{thompson}
W.~T.~Thompson,
``Transmission of elastic waves through a stratified solid medium,''
J.\ Appl.\ Phys.\ {\bf 21}, 89-93 (1950).

\bibitem{teitler-1970} 
S.~Teitler and B.~W.~Henvis, 
``Refraction in stratified, anisotropic media,'' 
J.\ Opt.\ Soc.\ Am.\ {\bf 60}, 830-834 (1970).

\bibitem{berreman-1972} 
D.~W.~Berreman, 
``Optics in stratified and anisotropic media: $4\times 4$-matrix formulation,''
J.\ Opt.\ Soc.\ Am.\ {\bf 62}, 502-510 (1972).

\bibitem{yeh} 
P.~Yeh, A.~Yariv, A., and C.-S.~Hong,
``Electromagnetic propagation in periodic stratified media. I. General theory,'' 
J.\ Opt.\ Soc.\ Am. {\bf 67}, 423-438 (1977).

\bibitem{abrahams-1980} 
E.~Abrahams and M.~J.~Stephen,
``Resistance fluctuations in disordered one-dimensional conductors,''
J.~Phys.~C: Solid St.\ Phys.\ {\bf 13}, L377-L381 (1980).

\bibitem{ardos-1982} 
P.~Erd\"os and R.~C.~,Herndon, 
``Theories of electrons in one-dimensional disordered systems,'' 
Adv.\ Phys.~{\bf 31}, 65-163 (1982).

\bibitem{pendry-1982} 
J.~B.~Pendry,  
``1D localisation and the symmetric group,''
J.~Phys.~C: Solid State Phys.\ {\bf 15} 4821-4834 (1982).

\bibitem{levesque} 
D.~Levesque and L.~Piche, 
``A robust transfer matrix formulation for the ultrasonic response of multilayered absorbing media,''
J.\ Acoust.\ Soc.\ Am.\ {\bf 92}, 452-467 (1992).

\bibitem{hosten} 
B.~Hosten and M.~Castaings, 
``Transfer matrix of multilayered absorbing and anisotropic media. Measurements and simulations of ultrasonic wave propagation through composite materials,''
J. Acoust. Soc. Am.\ {\bf 94}, 1488-1495 (1993).

\bibitem{sheng-1996} 
W.-D.\ Sheng and J.-B.\ Xia,  
``A transfer matrix approach to conductance in quantum waveguides,'' 
J.~Phys.: Condens. Matter {\bf 8} 3635-3645 (1996).

\bibitem{schubert-1996} 
M.~Schubert, 
``Polarization-dependent optical parameters of arbitrarily anisotropic homogeneous layered systems,''
Phys.\ Rev.~B {\bf 53}, 4265-4274 (1996).

\bibitem{wang-2001} 
L.~Wang and S.~I.~Rokhlin,  
``Stable reformulation of transfer matrix method for wave propagation in layered anisotropic media,''
Ultansonics {\bf 39}, 413-424 (2001).

\bibitem{wortmann-2002} 
D.~Wortmann, H.~Ishida, and S.~Bl\"ugel, 
``Ab initio Green-function formulation of the transfer matrix: Application to complex band structures,'' Phys.\ Rev.~B {\bf 65}, 165103 (2002).

\bibitem{katsidis-2002} 
C.~C.\ Katsidis and D.~I.\ Siapkas, 
``General transfer-matrix method for optical multilayer systems with coherent, partially coherent, and incoherent interference,'' 
App.\ Opt.~{\bf 41}, 3978-3987 (2002)

\bibitem{yeh-book} 
P.~Yeh, {\em Optical waves in layered media} (Wiley, Hoboken, NJ, 2005).

\bibitem{Hao-2008} 
J.~Hao and L.~Zhou, 
``Electromagnetic wave scattering by anisotropic metamaterials: Generalized $4\times 4$ transfer-matrix method,'' 
Phys.\ Rev.~B {\bf 77}, 094201 (2008).

\bibitem{li-2009}
H.~Li, L.~Wang, Z.~Lan, and Y.~Zheng, 
``Generalized transfer matrix theory of electronic transport through a graphene waveguide,'' 
Phys.\ Rev.~B {\bf 79}, 155429 (2009).

\bibitem{zhan-2013} 
T.~Zhan, X.~Shi, Y.~Dai, X.~Liu and J.~Zi,
``Transfer matrix method for optics in graphene layers,''
 J. Phys.: Condens. Matter {\bf 25} 215301 (2013).
 
\bibitem{jpa-2020b} 
F.~Loran and A. Mostafazadeh, 
``Transfer matrix for long-range potentials'' 
J.~Phys.~A: Math.\ Theor.\ {\bf 53}, 395303 (2020).

\bibitem{pereyray-1998a} 
P.~Pereyray, 
``Non-commutative polynomials and the transport properties in multichannel-multilayer systems,''
J~Phys.~A {\bf 31}, 4521-4531 (1998).

\bibitem{pereyray-2002} 
P.~Pereyray, 
``Theory of finite periodic systems: General expressions and various simple and illustrative examples,'' 
Phys.\  Rev.~B {\bf 65}, 205120 (2002).

\bibitem{pereyray-2005} 
P.~Pereyray, 
``Eigenvalues, eigenfunctions, and surface state in finite periodic systems,''
Ann.\ Phys.\ (N.Y.) {\bf 320}, 1-20 (2005).

\bibitem{Shukla-2005} 
P.~Shukla and I.~P.\ Batra, 
``Multichannel transport in a disordered medium under generic scattering conditions: A transfer-matrix approach,'' 
Phys.\ Rev.~B {\bf 71}, 235107 (2005).

\bibitem{anzaldo-meneses-2007} 
A.~Anzaldo-Meneses and P.~Pereyray, 
``Sylvester theorem and the multichannel transfer matrix method for arbitrary transverse potential profile inside a wave guide,''
Ann.\ Phys.\ (N.Y.) {\bf 322} 2114-2128 (2007).

\bibitem{pendry-1984} 
J.~B.~Pendry, 
``A transfer matrix approach to localisation in 3D,''
J.~Phys.~C: Solid State Phys.\ {\bf 17} 5317-5336 (1984).

\bibitem{pendry-1990a} 
J.~B.~Pendry,
``Transfer matrices and conductivity in two- and three-dimensional systems. I. Formalism,''
J.~Phys.: Condens.\ Matter {\bf 2}, 3273-3286 (1990).

\bibitem{pendry-1990b} 
J.~B.~Pendry, 
``Transfer matrices and conductivity in two- and three-dimensional systems. II. Application to localised and delocalised systems,'' 
J.~Phys.: Condens. Matter {\bf 2}, 3287-3301 (1990).

\bibitem{pendry-1994} 
J.~B.~Pendry,
``Photonic band structures,''
J.~Mod.\ Opt. {\bf 41}, 209-229 (1994).

\bibitem{mclean}
A.~S.~McLean and J.~B.~Pendry,
``A polarized transfer matrix for electromagnetic waves in structured media,'' 
J.~Mod.\ Opt. {\bf 41}, 1781-1802 (1994).

\bibitem{ward-1996} 
A.~J.~Ward and J.~B.~Pendry,
``Refraction and geometry in Maxwells equations,'' 
J.~Mod.\ Opt. {\bf 43}, 773-793 (1996).

\bibitem{pendry-1996} 
J.~B.~Pendry and P.~M.~Bell, 
``Transfer matrix techniques for electromagnetic waves,'' 
in {\em Photonic Band Gap Materials}, pp 203-228, edited by Soukoulis C.~M., NATO ASI Series, vol.\ 315 (Springer, Dordrecht, 1996).

\bibitem{pra-2016} 
F.~Loran and A.~Mostafazadeh,
``Transfer matrix formulation of scattering theory in two and three dimensions,'' 
Phys.\ Rev.~A \textbf{93}, 042707 (2016).

%\bibitem{prl-2009} 
%A.~Mostafazadeh,
%``Spectral singularities of complex scattering potentials and infinite reflection and transmission coefficients at real energies,'' 
%Phys.\ Rev.\ Lett.~\textbf{102}, 220402 (2009).

\bibitem{ap-2014} 
A.~Mostafazadeh,
``A Dynamical formulation of one-dimensional scattering theory and its applications in optics,''
Ann.\ Phys.\ (N.Y.)  \textbf{341}, 77 (2014).

\bibitem{pra-2014a} 
A.~Mostafazadeh,
``Transfer matrices as non-unitary S-matrices, multimode unidirectional invisibility, and perturbative inverse scattering,''
Phys.\ Rev.\ A {\bf 89}, 012709 (2014).

\bibitem{horsley-2015}
S.\ A.\ R. Horsley, M.\ Artoni and G.\ C.\ La Rocca,
``Spatial Kramers-Kronig relations and the reflection of waves,''
{Nature Photonics} \textbf{9}, 436-439 (2015).

\bibitem{longhi-2015}
S.\ Longhi,
``Wave reflection in dielectric media obeying spatial Kramers-Kronig relations,''
 {EPL} \textbf{112}, 64001 (2015).

\bibitem{horsley-longhi}
S.\ A.\ R. Horsley and S.\ Longhi,
``One-way invisibility in isotropic dielectric optical media,''
Amer.\ J.~Phys.~{\bf 85}, 439-446 (2017).

\bibitem{jiang-2017}
W.~Jiang, Y.~Ma, J.~Yuan, G.~Yin, W.~Wu, and S.~He,
``Deformable broadband metamaterial absorbers engineered with an analytical spatial Kramers-Kronig permittivity profile,''
Laser Photonics Rev.\ {\bf 11}, 1600253 (2017).

\bibitem{p171}
F.~Loran and A.~Mostafazadeh,
``Existence of the transfer matrix for a class of nonlocal potentials in two dimensions,'' in preparation.

\bibitem{Beauzamy}
B.~Beauzamy,
{\em Introduction to Operator Theory and Invariant Subspaces,}
(Elsevier Science Publications, Amsterdam, 1988).

\bibitem{kato} 
T.~Kato, 
{\em Perturbation Theory for Linear Operators} 
(Springer, Berlin, 1995).

\bibitem{Reed-Simon1}
M.~Reed and B.~Simon,
{\em Methods of Modern Mathematical Physics I: Functional Analysis}
(Academic Press, San Diego, 1980). 

\bibitem{ahues} 
M.~Ahues, A.~Largillier, and B.~V.~Limaye, 
{\em Spectral Computations for Bounded Operators}
(Chapman \& Hall/CRC, Roca Baton, 2001).

\bibitem{guven}
A.~G\"uven and O.~F.~Bandtlow,
``Quantitative spectral perturbation theory for compact operators on a Hilbert space,'' preprint arXiv: 2005.13891.

\end{thebibliography}
